\documentclass[aps,prc,reprint,showpacs,groupedaddress,onecolumn]{revtex4-1}

\usepackage{graphicx,color} 
\usepackage{dcolumn}  
\usepackage{bm}       
\usepackage{amsmath}
\usepackage{textcomp}
\usepackage{ifpdf}

\begin{document}

\newcommand {\nc} {\newcommand}

\newcommand{\vv}[1]{{$\bf {#1}$}}
\newcommand{\ul}[1]{\underline{#1}}
\newcommand{\vvm}[1]{{\bf {#1}}}

\nc {\IR} [1]{\textcolor{red}{#1}}
\nc {\IB} [1]{\textcolor{blue}{#1}}

\title{Separable Representation of Energy-Dependent  Optical Potentials}

\author{L.~Hlophe}
\email{lh421709@ohio.edu}
\author{Ch.~Elster}
\email{elster@ohio.edu}

\affiliation{Institute of Nuclear and Particle Physics,  and
Department of Physics and Astronomy,  Ohio University, Athens, OH 45701} 

\date{\today}

\begin{abstract}
\begin{description}
\item[Background] 
  One important ingredient for many applications of nuclear physics to astrophysics,
  nuclear energy,  and stockpile stewardship are cross sections for reactions of
neutrons with rare isotopes.
  Since direct measurements are often not feasible, indirect methods, e.g. (d,p)
reactions, should be used. 
  Those (d,p) reactions may be viewed as three-body reactions and described with
Faddeev techniques.

\item[Purpose] Faddeev equations in momentum space have a long tradition of
  utilizing separable interactions in order to arrive at sets of coupled integral
  equations in one variable. Optical potentials representing the effective
interactions in the neutron (proton) nucleus subsystem are usually non-Hermitian as well as
energy-dependent. Potential matrix elements as well as transition matrix elements
calculated with them must fulfill the reciprocity theorem. The purpose of this paper
is to introduce a separable, energy-dependent representation of complex, 
energy-dependent optical potentials that fulfill reciprocity exactly.

\item[Results] Starting from a separable, energy-independent representation of global
optical potentials based on a generalization of the Ernst-Shakin-Thaler (EST) scheme,
a further generalization is needed to take into account the energy dependence.
Applications to n$+^{48}$Ca,  n$+^{208}$Pb, and  p$+^{208}$Pb are investigated for
energies from 0 to 50~MeV with special emphasis on fulfilling reciprocity.

\item[Conclusions] We find that the energy-dependent separable representation 
of complex, energy-dependent phenomenological optical potentials fulfills reciprocity
exactly. In addition, taking into account the explicit energy dependence slightly
improves the description of the $S$ matrix elements.  
\end{description} 
\end{abstract}


\maketitle

\section{Introduction}
\label{intro}

Deuteron induced nuclear reactions are attractive from an experimental as well as theoretical
point of view to probe the structure of exotic nuclei. For example, carried out in inverse
kinematics, (d,p) or (d,n) reactions prove useful for extracting neutron or proton capture
rates for unstable nuclei of astrophysical relevance (see e.g.~\cite{Kozub:2012ka}). From a
theoretical perspective (d,p) and (d,n) reactions are attractive, since the
scattering problem may be viewed as an effective three-body
problem~\cite{ThompsonNunes}. One of the most challenging aspects of solving
the three-body problem for nuclear reactions is the repulsive Coulomb
interaction between the nucleus and the proton. While for very light
nuclei, exact calculations of (d,p) reactions based on momentum-space Faddeev equations in
the Alt-Grassberger-Sandhas (AGS)~\cite{ags} formulation can be carried
out~\cite{Deltuva:2009fp} by using a screening and renormalization
procedure~~\cite{Deltuva:2005wx,Deltuva:2005cc}, this technique leads to increasing technical
difficulties when moving to computing (d,p) reactions with
heavier nuclei~\cite{hites-proc}. Therefore, a new formulation of the Faddeev-AGS equations,
which does not rely on a screening procedure, was presented in
Ref.~\cite{Mukhamedzhanov:2012qv}. Here the Faddeev-AGS equations are cast in a
momentum-space  Coulomb-distorted partial-wave representation
instead of the plane-wave basis.  Thus all operators,  specifically the interactions
in the two-body subsystems must be evaluated in the Coulomb basis, which is a
nontrivial task (performed  recently for the neutron-nucleus
interaction~\cite{upadhyay:2014}).
The formulation of Ref.~\cite{Mukhamedzhanov:2012qv} requires the interactions in the
subsystems to be of separable form. 

Separable representations of the forces between constituents forming the subsystems in a
Faddeev approach have a long tradition, specifically when considering the nucleon-nucleon
(NN) interaction (see e.g.~\cite{Haidenbauer:1982if,Haidenbauer:1986zza,Entem:2001it}) or
meson-nucleon interactions~\cite{Ueda:1994ur,Gal:2011yp}. Here the underlying potentials are
Hermitian, and a scheme for deriving separable representations suggested by
Ernst-Shakin-Thaler~\cite{Ernst:1973zzb} (EST) is well suited, specifically when working in
momentum space. It has the nice property that the on-shell and half-off-shell transition
matrix elements of the separable representation are exact at predetermined energies, the
so-called EST support points.
However, when dealing with neutron-nucleus (nA) or proton-nucleus (pA)
phenomenological 
optical potentials, which are in general complex to account for absorptive channels that
are not explicitly treated, as
well as energy-dependent, extensions of the EST scheme have to be made. 

The generalization to non-Hermitian potentials which ensures that the potential as well as
the transition matrix elements fulfill the reciprocity theorem is given in
Ref.~\cite{Hlophe:2013xca}. Essential for this extension is the use of incoming as well as
outgoing scattering wave functions when setting up the EST scheme. For a separable
representation of pA optical potentials an EST construction has to be carried out in 
the basis of momentum space Coulomb functions instead of plane waves~\cite{Hlophe:2014soa}. 
Refs.~\cite{Hlophe:2013xca} and~\cite{Hlophe:2014soa} show that separable representations of 
phenomenological global optical potentials of Woods-Saxon type can readily be obtained
for light ($^{12}$C) as well as heavy ($^{208}$Pb) nuclei.

Strictly speaking, the EST representation of Ref.~\cite{Ernst:1973zzb} has the drawback that
its underlying assumptions rely on the energy independence of the original potential.
Unfortunately, today’s phenomenological global optical potentials are all energy-dependent. 
This drawback has been recognized by B.C. Pearce~\cite{Pearce:1987zz}, who showed how the
energy dependence of Hermitian potentials can be accommodated in the EST scheme. Those
suggestion were implemented for pion-nucleon interactions in Refs.~\cite{Pearce:1989re,Saito:2000bg}.

In Section~\ref{section_nA} we concentrate on nucleon-nucleus scattering
and show how the extension  of Pearce~\cite{Pearce:1987zz} to explicitly
accommodate energy dependence
in the EST scheme can be combined with our previous extension to non-Hermitian
potentials~\cite{Hlophe:2013xca}. Here we specifically give the additional momentum-space
terms  that need to be calculated.  In Section~\ref{result:nA} we show that those additional
terms mainly affect the off-shell behavior of partial-wave transition matrix elements, and
that only when taking into account the energy dependence reciprocity is fulfilled exactly.
We also study 
select partial-wave S-matrix elements for $n+^{48}$Ca and $n+^{208}$Pb,    
where we show that for on-shell quantities the explicit energy dependence has very little
effect.  In Section~\ref{section_pA} we show how this energy-dependent 
formulation of a separable representation can be employed for
proton-nucleus scattering, and present results for $p+^{208}$Pb, a case where the
Coulomb interaction is strong.
Since taking into account the energy dependence explicitly may considerably increase the 
computational effort when using this separable representation, we also
study the possibility of interpolating on the energy dependence of the optical potential.
Finally, we summarize our findings in Section~\ref{summary}.


\section{energy-dependent neutron-nucleus optical potentials}
\label{section_nA}

\subsection{Formal Considerations}
\label{formal:nA}

While the pioneering work by Ernst, Shakin and Thaler~\cite{Ernst:1973zzb}
constructed separable representations of Hermitian potentials, optical potentials
that describe the scattering of neutrons and protons from nuclei are in general
complex as well as energy-dependent. In Ref.~\cite{Hlophe:2013xca}
the EST scheme was extended to complex potentials. We briefly recall the most
important features, namely that a separable representation for a complex,
energy-independent potential
$U_l$ in a fixed partial wave of orbital angular momentum $l$ is given by~\cite{Hlophe:2013xca}
\begin{equation}
 u_l = \sum\limits_{ij} U_l|\psi_{l,i}^+ \rangle\lambda_{ij}^{(l)}\langle\psi_{l,j}^-|U_l, 
\label{eq:form1}
\end{equation}
where $|\psi_{l,i}^+\rangle$ is a solution of the Hamiltonian $H=H_0+U_l$ with outgoing
boundary conditions at energy $E_i$, and $|\psi_{l,i}^-\rangle$ is a
solution of the Hamiltonian $H=H_0+U_l^*$ with incoming boundary conditions. 
We refer to the energies $E_i$ as EST support points.
The free Hamiltonian  $H_0$ has 
eigenstates $|k_i\rangle$ with $k_i^2=2\mu E_i$, 
$\mu$ being  the reduced mass of the neutron-nucleus system. 
The matrix $\lambda_{ij}^{(l)}$ is constrained by the conditions
\begin{eqnarray}
 \delta_{kj}&=&\sum\limits_{i}\langle\psi_{l,k}^-|U_l|\psi_{l,i}^+\rangle\lambda_{ij}^{(l)} \cr
 \delta_{ik}&=&\sum\limits_{j}\lambda_{ij}^{(l)}\langle\psi_{l,j}^-|U_l|\psi_{l,k}^+\rangle,
 \label{eq:form1b}
 \end{eqnarray}
where the subscript $i=1 \dots N$ indicates the rank of the separable potential.
The two constraints of Eq.~(\ref{eq:form1b})  on $\lambda_{ij}^{(l)}$
 ensure that at the EST support points $E_i$, both the original $U$
and the separable potential $u$, yield identical
wavefunctions or half-shell $t$ matrices. 
The corresponding separable
$t$ matrix takes the form
\begin{equation}
 t_l(E) = \sum\limits_{ij}U_l|\psi_{l,i}^+\rangle\tau^{(l)}_{ij}(E)
  \langle\psi_{l,j}^-|U_l
\label{eq:form2}
\end{equation}
with
\begin{equation}
  \left(\tau^{(l)}_{ij}(E)\right)^{-1}= \langle\psi_{l,i}^-|U_l-U_lg_0(E)U_l|\psi_{l,j}^+\rangle.
  \label{eq:form2b}
\end{equation}
Here $g_0(E) = (E -H_0 +i \varepsilon)^{-1}$ is the free propagator. The
form factors are given as
half-shell $t$-matrices
\begin{equation}
T_l(E_i)|k_i\rangle\equiv U_l|\psi_{l,i}^+\rangle,
\label{eq:form2c}
\end{equation}
and are obtained through solving a momentum space Lippmann-Schwinger (LS) equation.
 
Given a time reversal operator $\mathcal{K}$, the optical potential $U$ satisfies
\begin{equation}
\mathcal{K}U\mathcal{K}^{-1}=U^\dagger,
\label{eq:form3}
\end{equation}
 so that the corresponding $t$-matrix fulfills reciprocity. We therefore require that 
 its EST separable representation $u$ preserves this property. A proof
 that the separable potential defined in Eqs.~(\ref{eq:form1}) and~(\ref{eq:form2}) obeys the relation 
$\mathcal{K}u\mathcal{K}^{-1}=u^\dagger$
 is provided in Ref.~\cite{Hlophe:2013xca} for the rank-1 case. Generalization 
 of the proof to a higher rank requires that the matrix $\lambda_{ij}^{(l)}$ be symmetric
in the indices $(i,j)$.  
If the potential is energy-independent this symmetry of  $\lambda_{ij}^{(l)}$ is 
 evident when examining the constraints of Eq.~(\ref{eq:form1b}). 
 
However,  when applying the same formulation to an energy-dependent potential $U(E)$, 
one obtains
\begin{equation}
  u_l = \sum\limits_{ij}U_l(E_i)|\psi_{l,i}^+\rangle\lambda^{(l)}_{ij}\langle
\psi_{l,j}^-|U_l(E_j),
 \label{eq:form4}
\end{equation}
with the constraints
\begin{eqnarray}
\delta_{kj}&=&\sum\limits_{i}\langle\psi_{l,k}^-|U_l(E_i)|\psi_{l,i}^+\rangle
 \lambda^{(l)}_{ij} \cr
\delta_{ik}&=&\sum\limits_{j}\lambda^{(l)}_{ij}\langle\psi_{l,j}^-|U_l(E_j)|
\psi_{l,k}^+\rangle.
 \label{eq:form4b}
 \end{eqnarray}
Omitting the partial wave index $l$  the two constraints on $\lambda$ 
can be written in matrix form
\begin{equation}
\mathcal{U}^t \; \lambda = {\bf 1} =\lambda \;  \mathcal{U},
\label{form5}
\end{equation}
with
\begin{equation}
\mathcal{U}_{ij} = \langle\psi_i^-|U(E_i)|\psi_j^+\rangle.
   \label{form5b}
\end{equation}
For a separable potential of rank $N>1$ the matrix $\mathcal{U}_{ij}$ is not symmetric
in the indices $i$ and $j$,  
which leads to an asymmetric matrix $\lambda$ and thus a $t$ matrix which violates 
reciprocity. 
Therefore, a different approach must be taken in order to  construct separable
representations for energy-dependent potentials. Here we note that although 
the potential $u$ contains some of the energy dependence of $U(E)$
through the form factors, it has no explicit energy dependence.
Henceforth we will refer to this separable construction as the energy-independent
EST representation.

A separable expansion for energy-dependent Hermitian potentials was suggested by B.C.
Pearce~\cite{Pearce:1987zz}. This suggestion can also be applied to complex
potentials by using the insights already gained in~\cite{Hlophe:2013xca}.
In analogy, we define the EST separable
representation for complex energy-dependent potentials (eEST) by allowing
an explicit energy dependence of the coupling matrix elements $\lambda_{ij}$.
\begin{eqnarray}
u(E) = \sum\limits_{ij}U(E_i)|\psi_i^+\rangle\lambda_{ij}(E)\langle\psi_j^-|U(E_j),
\label{eq:form9}
\end{eqnarray}
where the partial wave index $l$ has been omitted for simplicity. In order
to obtain a constraint on the matrix $\lambda(E)$, we require that the matrix
elements of the potential $U(E)$ and its separable form $u(E)$ between the 
states $|\psi_i^+\rangle$ be the same at 
all energies $E$. This condition ensures that the potentials
$U(E)$ and $u(E)$ yield identical wavefunctions at the EST support points,
just like in the energy-independent EST scheme.
The constraints on $\lambda_{ij}(E)$ become
\begin{eqnarray}
\langle\psi_m^-|U(E)|\psi_n^+\rangle
  &=&\langle\psi_m^-|u(E)|\psi_n^+\rangle\cr
  &=&\sum\limits_{i}\langle\psi_m^-|U(E_i)|\psi_i^+\rangle\lambda_{ij}(E)
  \langle\psi_j^-|U(E_j)|\psi_n^+\rangle.\label{eq:form10}
 \end{eqnarray}
 The corresponding separable $t$-matrix then takes the form 
\begin{eqnarray}
 t(E) = \sum\limits_{ij}U(E_i)|\psi_i^+\rangle\tau_{ij}(E)\langle\psi_j^-|U(E_j).
\label{eq:form11}
\end{eqnarray}
Substituting Eqs.~(\ref{eq:form9})$-$(\ref{eq:form11}) into the LS equation leads to
the constraint on the matrix $\tau(E)$ such that
\begin{equation}
 R(E)\cdot\tau(E) \equiv \mathcal{M}(E),
\label{eq:form12}
\end{equation}
where
\begin{equation}
 R_{ij}(E) = \langle\psi_i^-|U(E_i)|\psi_j^+\rangle-\sum\limits_{n}\mathcal{M}_{in}(E)
\langle\psi_n^-|U(E_n)\; g_0(E)\; U(E_j)|\psi_j^+\rangle,
\label{eq:form12b}
  \end{equation}
with
\begin{eqnarray}
\mathcal{M}_{in}(E) \equiv[\mathcal{U}^e(E)\cdot \mathcal{U}^{-1}]_{in}, 
\label{eq:form12c}
\end{eqnarray}
where the matrix elements of $\mathcal{U}$ are defined in Eq.~(\ref{form5b}). 
We want to point out that, for energy-independent potentials, the matrix
$\mathcal{M}(E)$ simply is the unit matrix.
For further evaluating these matrix elements
 in momentum space, we express 
 
\begin{equation}
\mathcal{U}^e_{ij}(E) \equiv \langle\psi_i^-|U(E)|\psi_j^+\rangle,
 \label{eq:form13a}
\end{equation}
in terms of the potential and the half-shell $t$ matrix so that
 \begin{eqnarray}
\mathcal{U}^e_{ij}(E) 
 &=&\langle k_i|U(E)|k_j\rangle+\langle\psi_i^-|U(E_i) \; g_0(E_i) \; U(E)|k_j\rangle
    +\langle k_i|U(E) \; g_0(E_j) \; U(E_j)|\psi_j^+\rangle\cr
    &&+\langle \psi_i^-|U(E_i) \; g_0(E_i) \; U(E)g_0(E_j)U(E_j)|\psi_j^+\rangle,\cr
   &=&\langle k_i|U(E)|k_j\rangle+\langle k_i|T(E_i) \; g_0(E_i) \; U(E)|k_j\rangle
    +\langle k_i|U(E) \; g_0(E_j) \; T(E_j)|k_j \rangle\cr
    & +&\langle k_i|T(E_i) \; g_0(E_i) \; U(E) \; g_0(E_j) \; T(E_j)|k_j \rangle .
 \label{eq:form13}
\end{eqnarray}
Inserting a complete set of momentum eigenstates leads to the explicit expression 
for $\mathcal{U}^e_{ij}(E)$,
\begin{eqnarray}
\mathcal{U}^e_{ij}(E) 
&=&U(k_i,k_j,E)+\int\limits_0^\infty dp p^2 \;T(p,k_i;E_i)\;g_0(E_i,p)\;U(p,k_j,E) \cr
    &+&\int\limits_0^\infty dp p^2\; U(k_i,p,E)\;g_0(E_j,p)\;T(p,k_j;E_j)\cr
    &+&\int\limits_0^\infty dp p^2 \int\limits_0^\infty dp' p'^2\; 
    T(p,k_i;E_i)\;g_0(E_i,p)\; U(p,p',E) \;g_0(E_j,p')\;T(p',k_j;E_j),
 \label{eq:esep8d0}
\end{eqnarray}
where $g_0(E,p)=[E-p^2/2\mu+i\varepsilon]^{-1}$.  
For the evaluation of $\mathcal{U}^e_{ij}(E)$ for all energies $E$ within the relevant 
energy regime, we need the form factors $T(p',k_j;E_j)$ at the specified EST
support points as well as the matrix elements of the potential $U(p',p,E)$ at
all energies.  


\subsection{Energy-independent EST Scheme versus eEST Separable Representation}
\label{result:nA}

 For studying the properties of the energy-dependent separable 
 representation eEST, we consider elastic scattering of neutrons 
 off $^{48}$Ca and $^{208}$Pb from 0 to 50~MeV.
 We employ the Chapel Hill (CH89) phenomenological global optical
 potential~\cite{Varner:1991zz} in all calculations. First, we 
 investigate the symmetry properties of the  off-shell partial wave
 $t$ matrix $t_{l}^{j}(k',k;E)$ in the eEST separable 
 representation and contrast them with the ones obtained via the energy 
 independent EST scheme. To do so,
 we adopt the same energy support points for the both EST separable 
 representations.
 Fig.~\ref{fig1} shows the off-shell $t$ matrix $t_{l=6}^{j=13/2}(k',k;E)$ 
 as function of the off-shell momenta $k$ and $k'$ for the $n+^{48}$Ca system
 at $E_{lab}=16$~MeV. 
 The real and imaginary parts of the off-shell $t$ matrix
 evaluated with the CH89 optical potential are shown in panels (a) and (d).
 The energy-independent EST calculation is given in panels (b) and (e),
 while the eEST separable representation is depicted in in panels (c) and (f).
 We observe that the structure of the off-shell separable $t$ matrix appears
 to be the same for both the energy-dependent and energy-independent representations.
 The high momentum components which are visible in the CH89 off-shell 
 $t$ matrix are projected out by both separable representations. This
 is a general feature of the EST separable representation. As further example, 
 we consider neutron scattering off the much heavier
 $^{208}$Pb nucleus as depicted in Fig.~\ref{fig2} for the $l=0$ partial
 wave. In both figures the separable off-shell $t$-matrices appear to be 
 symmetric around the $k=k'$ line. However we know from the formal considerations
 in the previous section that the energy-independent EST scheme does
 not fully satisfy reciprocity and therefore should yield an asymmetric
 off-shell $t$ matrix in $k$ and $k'$. In order to carry out a more careful
 analysis of the symmetry properties of the $t$-matrix we define an asymmetry
 
  \begin{eqnarray}
  \Delta t_l^j(k',k;E)=\frac{\left|t_l^j(k',k;E)-t_l^j(k,k';E)\right|}
  {\frac{1}{2}\left|t_l^j(k',k;E)+t_l^j(k,k';E)\right|},
  \label{eq:res1}
 \end{eqnarray}
 representing the relative difference between the off-shell $t$ matrix and its
 transpose. For a completely symmetric off-shell $t$ matrix this asymmetry
 should be exactly zero. In Fig.~\ref{fig3} we show the asymmetry
 $\Delta t_{l=6}^{j=13/2}(k',k;E)$ for $n+^{48}$Ca scattering. Panels (a) and (b)
 show the asymmetry for the energy-independent
 EST separable representation at $E_{lab}=16\;\text{and}\;40$~MeV. Panels (c) and (d)
 depict the asymmetry for the eEST separable representation at $E_{lab}=16\;\text{and}
 \;40$~MeV. For the energy-independent EST representation the asymmetry is either zero or very 
 small close to the $k=k'$ axis and at small momenta. However, away from this region
 it can become quite large. For the eEST separable representation the asymmetry
 is exactly zero everywhere as expected. This shows that in order to exactly fulfill
 reciprocity the eEST separable representation must be employed. 

 So far we only considered off-shell properties of the eEST separable
 representation. The next question is whether there is an on-shell difference
 between the energy-dependent and energy-independent schemes. As a measure of the
 quality of the eEST separable representation, we define the relative error of the
 real part of the $S$ matrix as
 
  \begin{equation}
  \text{relative\;error}=\left|\frac{{Re\;S_l^j(E)}^{orig}-{Re\;S_l^j(E)}^{sep}}
  {{S_l^j(E)}^{orig}}\right|,
  \label{eq:res2}  
 \end{equation}
where ${S_l^j(E)}^{orig}$ is the partial wave $S$ matrix calculated from the CH89 potential and
${S_l^j(E)}^{sep}$ the one obtained from the separable representation. The real part
of the $n+^{48}$Ca $S$ matrix for $l=6$ and $j=13/2$ together with the corresponding relative
error is depicted in panels (a) and (b) of Fig.~\ref{fig4}. The $S$ matrix obtained from the CH89 
phenomenological optical potential is shown by the
dash-dotted line while the energy-independent and eEST separable representations are depicted by
dashed and solid lines. The relative error is indicated by upward triangles for the
EST separable representation and by circles for the eEST scheme. There
is good agreement between the CH89 $S$ matrix and both separable representations. 
However the eEST separable representation describes the $S$ matrix slightly better than
its energy-independent counterpart since it incorporates more of the 
energy dependence of the original potential. This is visible for energies around $E_{lab}=37$~MeV,
where the relative error is dominated by the separable approximation. However, the energy-independent 
EST representation can always be improved by adding an extra
support point, i.e. increasing the rank. This means that the observations made in
Ref.~\cite{Hlophe:2013xca}  concerning on shell properties of the energy-independent 
EST representation apply to the eEST separable representation as well. The main reason
for adopting the eEST separable representation is that it yields exact reciprocity.


\section{Application to Proton-Nucleus Optical Potentials}
\label{section_pA}

\subsection{Formal Considerations}
\label{formal:pA}

The proton-nucleus potential consists of the point Coulomb force, $V^c$, together with
a short-ranged nuclear as well as a short-ranged  Coulomb interaction representing the charge
distribution of the nucleus, which we refer to as $U^s(E)$. While the point Coulomb 
potential has a simple analytical form, an optical potential is employed
to model the short-range nuclear potential. The extension of the energy-independent EST
separable representation to proton-nucleus optical potentials was carried out 
in Ref.~\cite{Hlophe:2014soa}. In that work it was shown that the 
form factors of the separable representation are solutions of the LS equation in the Coulomb basis,
and that  they are obtained using methods introduced in 
Refs.~\cite{Elster:1993dv,Chinn:1991jb}. It was also demonstrated that the extension of the energy-independent 
EST separable representation scheme to proton-nucleus scattering involves two steps. First, the
nuclear wavefunctions $|\psi_{l,i}^{(+)}\rangle$ are replaced by Coulomb-distorted nuclear wavefunctions
$|\psi_{l,i}^{sc~(+)}\rangle$. Second, the free resolvent $g_0(E)$ is replaced by the Coulomb Green's
function, $g_c(E) =(E-H_0 -V^c +i\varepsilon)^{-1}$.
As demonstrated in Section~\ref{section_nA}, in order
to fulfill reciprocity, an energy-dependent separable representation must be adopted. This is
accomplished by generalizing the eEST scheme to proton-nucleus scattering analogous 
to the extension of the energy-independent EST scheme presented in Ref.~\cite{Hlophe:2014soa}.
Thus applying the two steps outlined above to the eEST scheme
yields the separable Coulomb-distorted nuclear $t$-matrix 
\begin{equation}
 t^{sc}_l(E) =  \sum_{i,j} U_l^s(E_i)|\psi_{l,j}^{sc~(+)}\rangle \;\tau^{c,\;l}_{ij}(E)\;
 \langle \psi_{l,j}^{sc~(+)}|U_l^s(E_j) .
 \label{eq:pform1}
\end{equation}
Here $|\psi_{l,i}^{sc~(+)}\rangle $ are solutions corresponding to $U^s_l(E_i)$ in 
the Coulomb basis with outgoing boundary conditions, and $|\psi_{l,i}^{sc~(-)}\rangle$ are solutions
corresponding to $(U^s_l)^*(E_i)$  with incoming
boundary conditions. 
Upon suppressing the index $l$ we obtain a constraint similar to Eq.~(\ref{eq:form12}),
\begin{eqnarray}
R^{c}(E)\cdot\tau^{c}(E)
=\mathcal{M}^{c}_{ij}(E),
\label{pform2}
\end{eqnarray}
with the matrix elements of $R^{c}(E)$ satisfying
\begin{eqnarray}
 R_{ij}^{c}(E)=\langle\psi_{i}^{sc~(-)}|U^s(E_i)|\psi_{j}^{sc~(+)}\rangle -
 \sum_i\mathcal{M}^{c}_{in}(E)\langle\psi_{n}^{sc~(-)}|U^s(E_n)g_c(E) U^s(E_j)|\psi_{j}^{sc~(+)}\rangle.
\label{eq:pform3}
\end{eqnarray}
 The matrix $\mathcal{M}^{c}(E)$ is the Coulomb distorted counterpart
 of $\mathcal{M}(E)$ of Eq.~(\ref{eq:form12c}), and is defined
 as 
\begin{equation}
\mathcal{M}^c_{in}(E) = \left[\mathcal{U}^{e,sc}(E)\cdot ({\mathcal{U}^{sc}})^{-1}\right]_{in},
\label{eq:pform4}
\end{equation}
with
\begin{eqnarray} 
\mathcal{U}^{sc}_{ij}&\equiv& \langle\psi_{i}^{sc~(-)}|U^s(E_i)|\psi_{j}^{sc~(+)}\rangle,\cr
\mathcal{U}^{e,sc}_{ij}(E)&\equiv&\langle \psi_{k_i}^{sc~(-)} | U^s(E) | \psi_{k_j}^{sc~(+)} \rangle.
\label{eq:pform4b}
\end{eqnarray}
 If the potential is energy-independent the matrix $\mathcal{M}^c(E)$ becomes a unit matrix just like
 $\mathcal{M}(E)$.

For evaluating  $\mathcal{U}^{e,sc}_{ij}(E)$ on can proceed 
analogously to Eq.~(\ref{eq:form13}) and finally arrive at an expression similar to
Eq.~(\ref{eq:esep8d0}), namely
 \begin{eqnarray}
\mathcal{U}^{e,sc}_{ij}(E) 
&=&U^{sc}(k_i,k_j,E)+\int\limits_0^\infty dp p^2\; T^{sc}(p,k_i;E_i)g_c(E_i,p)U^{sc}(p,k_j,E) \cr
    &+&\int\limits_0^\infty dp p^2\; U^{sc}(k_i,p,E)g_c(E_j,p)T^{sc}(p,k_j;E_j)\cr\cr
    &+&\int\limits_0^\infty dp p^2\; \int\limits_0^\infty dp' p'^2\;
    T^{sc}(p,k_i;E_i)g_c(E_i,p) U^{sc}(p,p',E) g_c(E_j,p')T^{sc}(p',k_j;E_j).
 \label{eq:pform5}
\end{eqnarray}
Since all matrix elements are evaluated in the Coulomb basis, the Coulomb Green's function has the same
form as the free Green's function in the calculations.
The matrix elements of the short-ranged potential in the basis of 
Coulomb scattering states $U^{sc}(k_i,k_j,E)
\equiv\langle \phi_{k_i}^{c~(+)} |U^s| \phi_{k_j}^{c~(+)}\rangle$
are calculated in the same fashion as the ones in  Eq.~(9) of Ref.~\cite{Hlophe:2014soa}. The Coulomb
distorted short-ranged half-shell $t$ matrix $T^{sc}(p,k_i,E_i)$ is then evaluated using
Eq.~(6) of the same reference.

\subsection{Energy-independent EST Scheme versus eEST Separable Representation}
\label{results:pA} 

\begin{figure}
\begin{center}
    \includegraphics[scale=.45]{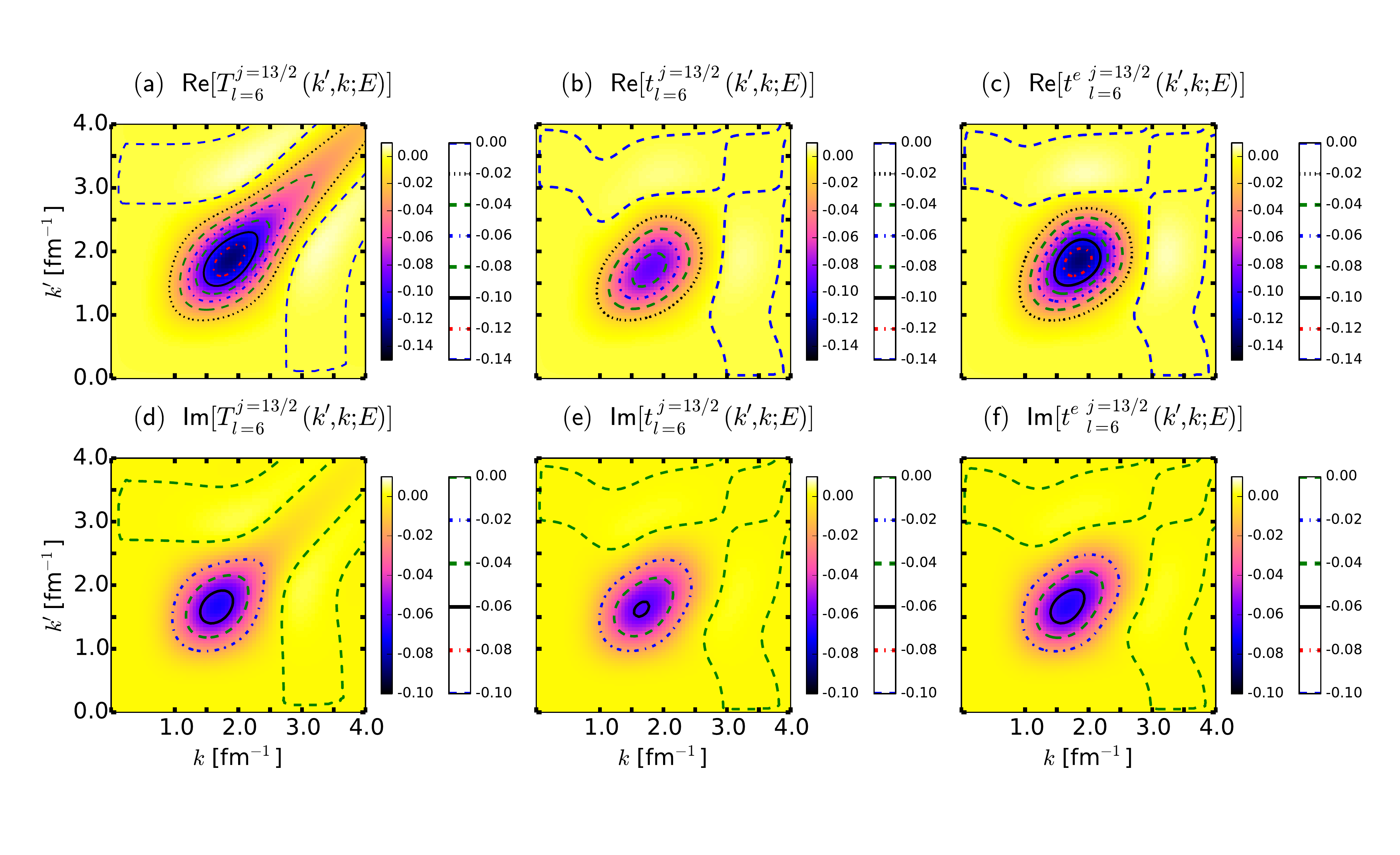}
\caption{(Color online)
   The $l = 6, j = 13/2$ partial wave off-shell $t$-matrix 
   elements, $t_6 (k' , k; E)$ in units fm$^2$ for the $n+^{48}$Ca system computed 
   at $E_{lab} =$ 16~MeV as function of the off-shell momenta $k'$ and $k$. This
   energy corresponds to an on shell momentum of  0.86 fm$^{-1}$. The real
   and imaginary parts of the  off-shell $t$ matrix calculated from the  CH89~\cite{Varner:1991zz}
   phenomenological optical potential are shown in panels (a) and (d).
   The real and imaginary parts of the $t$ matrix calculated from its energy-independent EST separable representation
   are shown in panels  (b) and (e), while panels (c) and (f) 
   depict the energy-dependent eEST separable representations. The support 
   points for the separable representation are at $E_{lab}$=16, 29, and 47~MeV.  
\label{fig1}
}
\end{center}
\end{figure}

\begin{figure}
\begin{center}
\includegraphics[scale=.45]{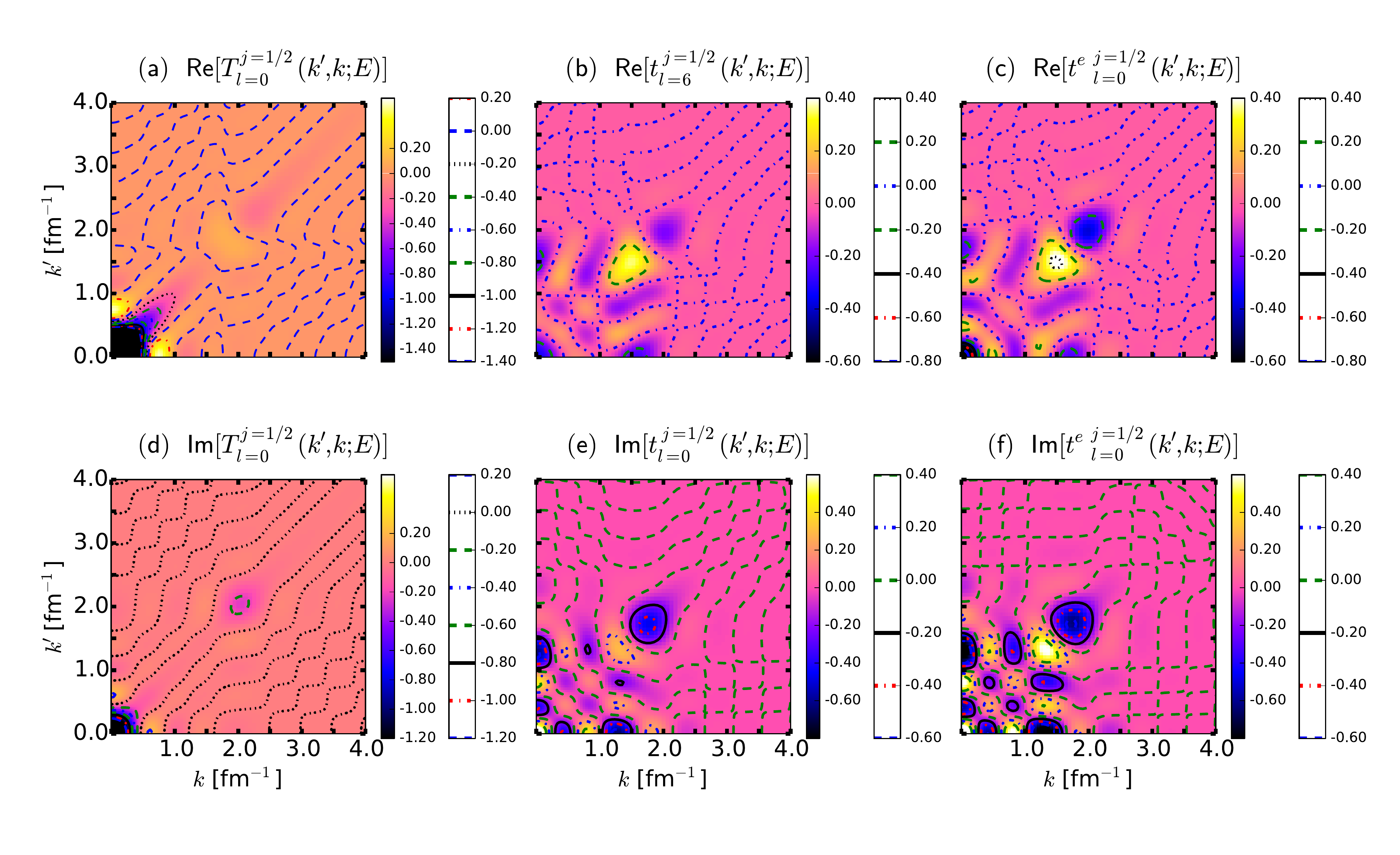}
\caption{(Color online)
 Same as FIG.~\ref{fig1} but for the $l=0,\; j=1/2$ partial wave of the $n+^{208}$Pb
 system at 21~MeV corresponding to an on shell momentum of 1.00 fm$^{-1}$. The support 
points for the separable representation are at $E_{lab}$=5, 11, 15, 21, and 47~MeV.
}
\label{fig2}

\end{center}
\end{figure}

\begin{figure}[ht]
\centering
\includegraphics[scale=0.45]{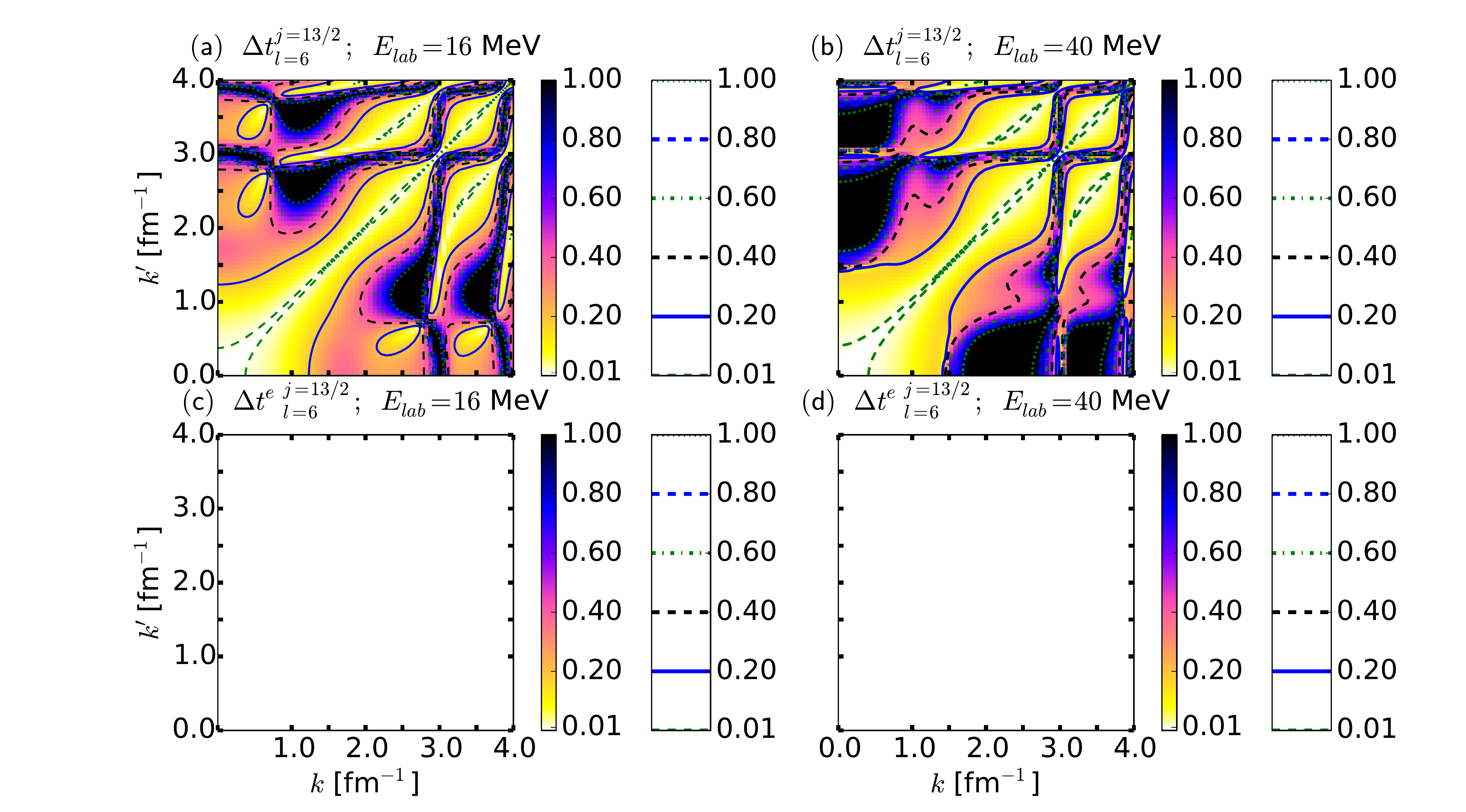}
\caption{(color online) The  asymmetry $\Delta t_{l=6}^{j=13/2} (k' , k; E)$
computed at $E_{lab} =$ 16 and 40~MeV as function of the off-shell momenta $k'$ and $k$
for the $n+^{48}$Ca system. 
Panels (a) and (c) on the left hand side show the asymmetry of the energy-independent
and energy-dependent separable representation of the $t$ matrix obtained from the CH89 
phenomenological optical potential at $E_{lab} =$ 16~MeV. Panels (b) and (d) on the right
hand side depict the asymmetry for the energy-independent and energy-dependent EST 
   separable representation at 40~MeV. The support points are $E_{lab}$= 16, 29, and 47~MeV.
   The on shell momenta are 0.86 fm$^{-1}$ and 1.36 fm$^{-1}$.
     }
   \label{fig3}
 \end{figure}

\begin{figure}[ht]
\centering
\includegraphics[scale=0.4]{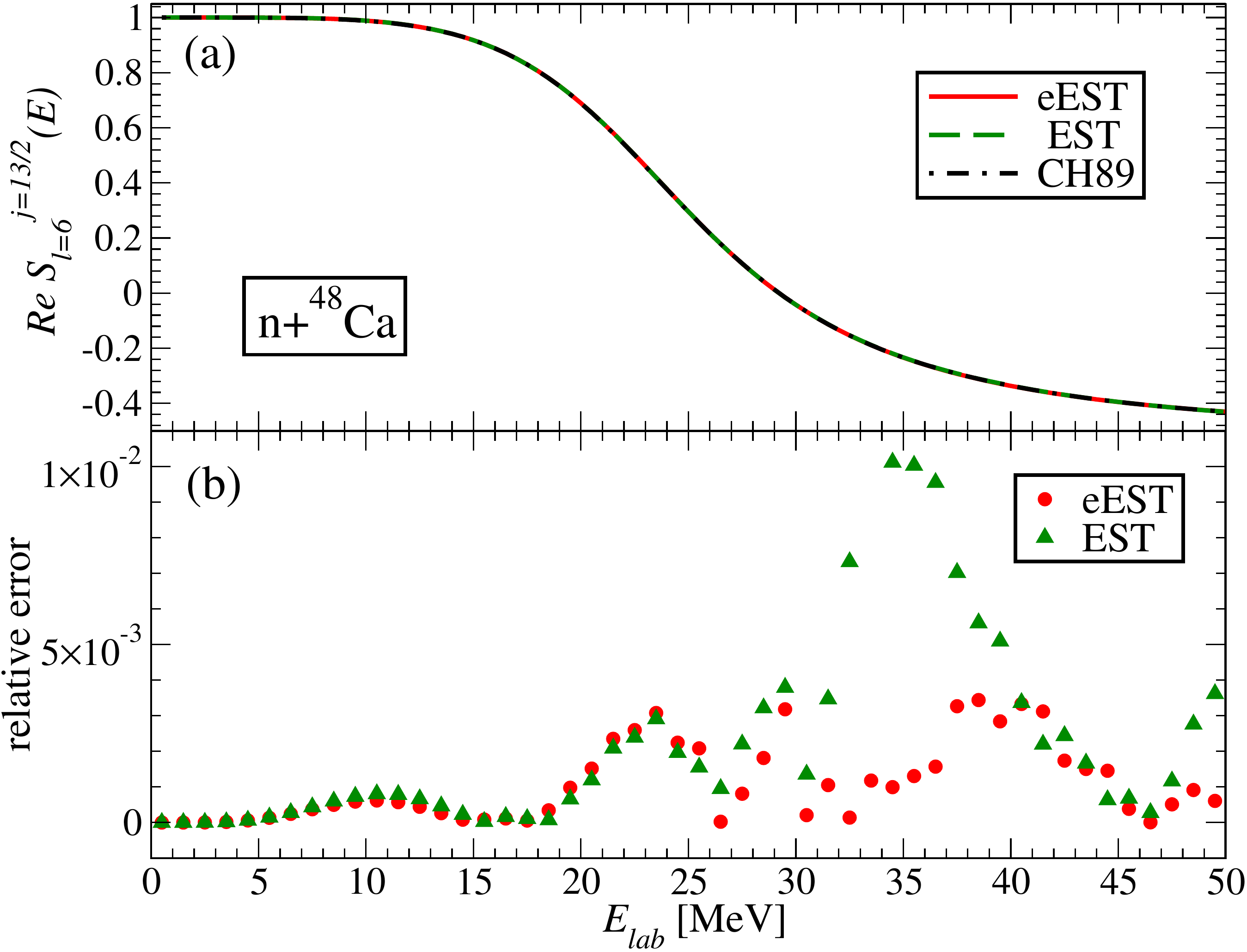}
\caption{(color online) The $S$ matrix elements $S_{l=6}^{j=13/2}(E)$ for elastic scattering
of neutrons from $^{48}$Ca in $l=6,\;j=13/2$ partial wave as function of the 
laboratory energy.
The top panel (a) shows the real part of the $S$ matrix while the bottom panel (b) 
gives the relative error of the separable representations as defined in Eq.~(\ref{eq:res1}).
The $S$ matrix calculated with the CH89 optical potential is represented by a (black) dash-dotted line,
the (red) solid line shows the energy-dependent EST (eEST) separable representation of the $S$ matrix,
while the energy-independent EST separable representation is indicated by a (green) dashed line.  The relative error is depicted by (red) circles for the eEST separable representation and by (green) upward
   triangles for the energy-independent EST construction. The EST support points are at $E_{lab}= 16,\;29,\text{~and~} 47$~MeV.   
}
\label{fig4}
\end{figure}

\begin{figure}
\begin{center}
\includegraphics[scale=.45]{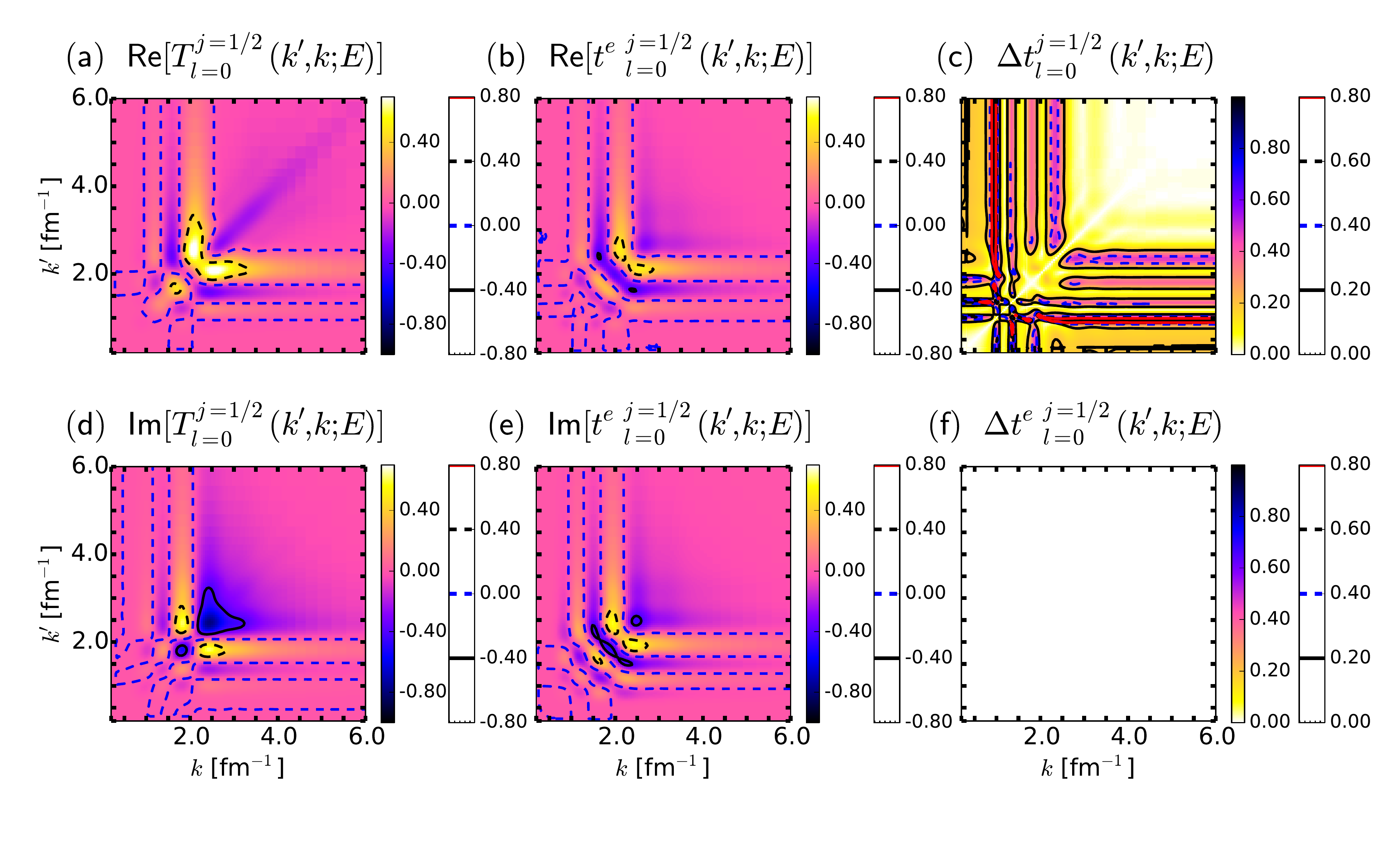}
\caption{(Color online)
   The $l = 0, j = 1/2$ partial wave off-shell $t$-matrix 
   elements, $t_0 (k' , k; E)$ in units fm$^2$ for the $p+^{208}$Pb system computed 
   at $E_{lab} =$ 21~MeV as function of the off-shell momenta $k'$ and $k$. This
   energy corresponds to an on shell momentum of  1.00 fm$^{-1}$. The real
   and imaginary parts of the  off-shell $t$ matrix calculated from the  
   CH89 phenomenological optical potential are shown in panels (a) and (d).
   The real and imaginary parts of its
   eEST separable representation are depicted in panels  (b) and (e). 
   Panels (c) and (f) depict the asymmetry for the energy-independent EST
   and eEST separable representations. The support points for the separable 
 representation are at $E_{lab}$=5, 11,
   21, 36, and 47~MeV.  
\label{fig5}
}
\end{center}
\end{figure}

In this section the generalization of the eEST separable representation 
is applied to elastic scattering of
protons off $^{208}$Pb.
The short-range interaction $U^s(E)$ is comprised of the CH89
global optical potential~\cite{Varner:1991zz} and a short-ranged Coulomb potential,
representing the charge distribution of the $^{208}$Pb nucleus as used for the
calculations of Ref.~\cite{Hlophe:2014soa}. As in Section~\ref{result:nA}, we first
concentrate on the off-shell $t$ matrices to verify that the  energy
dependent eEST representation for charged particles fulfills reciprocity exactly.
Since the Coulomb distortion of plane wave states
is most pronounced in low partial waves,  the 
the s-wave off-shell $t$ matrix is examined. Panels (a) and (d) of Fig.~\ref{fig5}
depict the real and imaginary parts of the off-shell $t$ matrix for the $p+^{208}$Pb
system calculated with the CH89 potential. 
Comparing those panels to the corresponding ones in Fig.~\ref{fig2} shows that the
attractive part of the $n+^{208}$Pb $t$ matrix at low momenta $k$ and $k'$ is absent in
the $p+^{208}$Pb $t$ matrix, indicating that the Coulomb interaction dominates here.
Since we already showed in Figs.~\ref{fig1} and~\ref{fig2} that there is no visual
difference in the off-shell $t$ matrices when comparing the energy-independent and
energy-dependent separable representation, we only show the eEST separable
representation in Fig.~\ref{fig5} in panels (b) and (d). As already observed for the
neutron off-shell $t$ matrices of Figs.~\ref{fig1} and~\ref{fig2}, the separable
representation projects out the high momentum components.

The asymmetry calculated according to Eq.~(\ref{eq:res1}) for the energy-independent separable representation
is illustrated in panel (c) while panel (f) depicts the asymmetry for the eEST
separable representation. 
As was the case for the $n+^{208}$Pb and $n+^{48}$Ca off-shell $t$ matrices, the eEST
representation generalized for proton scattering is completely symmetric in the
momenta $k$ and $k'$, leading to a zero asymmetry. This is not the case for the energy
independent EST generalization of Ref.~\cite{Hlophe:2014soa}, which is shown in panel~(c).
This demonstrates that  also the eEST representation
of proton-nucleus optical potentials fulfills reciprocity exactly. 

\begin{figure}[ht]
\centering
\includegraphics[scale=0.4]{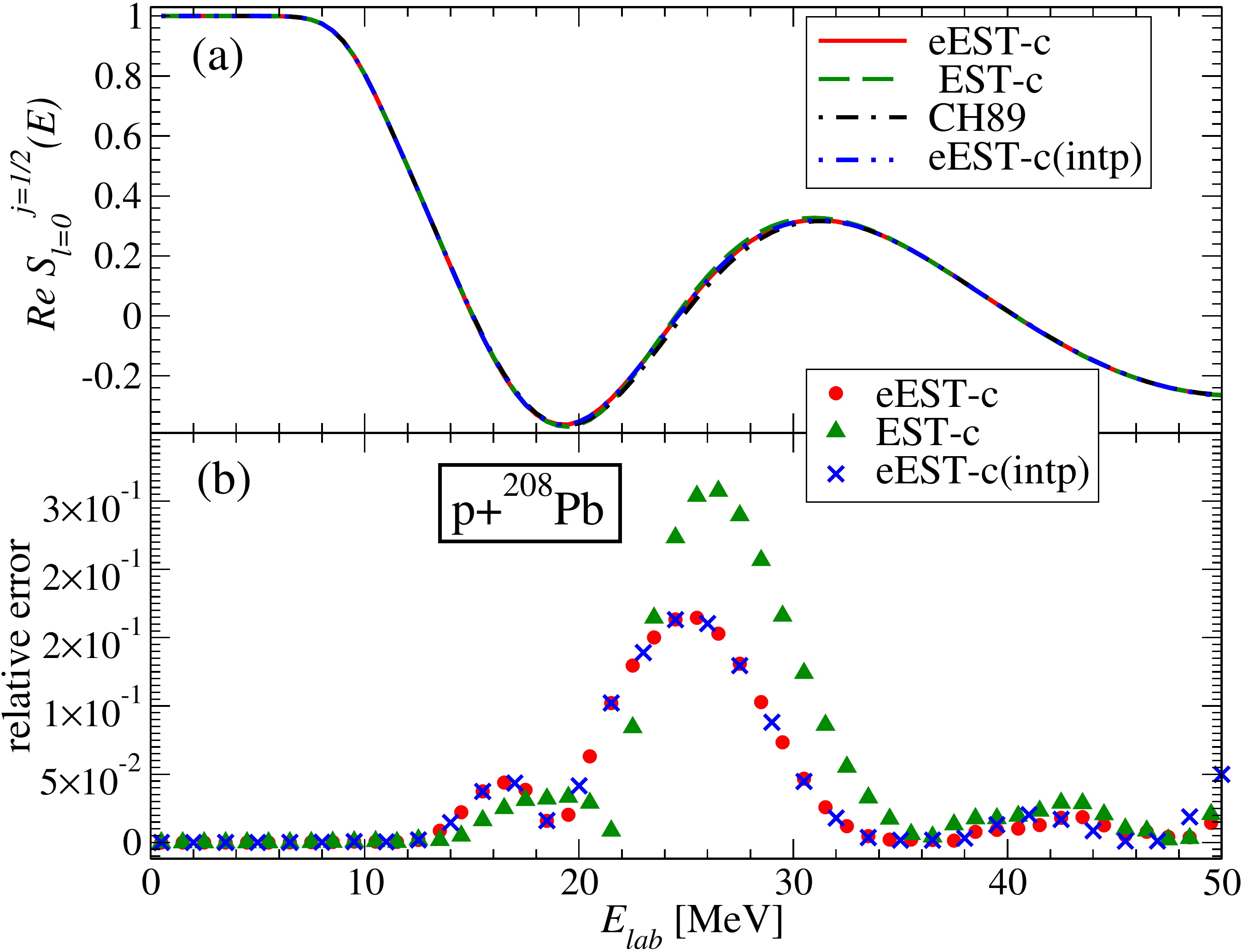}
\caption{(color online) The  $S$ matrix elements $S_{l=0}^{j=1/2}(E)$ for elastic
scattering of protons from $^{208}$Pb  in the $l=0$, $j=1/2$ partial wave as function 
 of the   laboratory energy.
The top panel (a) shows the real part of the $S$ matrix while the bottom panel (b) 
shows the relative error of the separable representations as defined in Eq.~(\ref{eq:res2}).
The $S$ matrix calculated from the CH89 phenomenological optical
potential~\cite{Varner:1991zz} is represented by a (black) dash-dotted line.
   The (red) solid line shows the energy-dependent EST (eEST-c) separable representation of the $S$ matrix.
   The energy-independent EST separable representation (EST-c)  is indicated by a (green) dashed line. The relative error
   is depicted by (red) circles for the eEST separable representation and by (green) upper
   triangles for the energy-independent EST construction. The EST support points for this case are $E_{lab}=
   5,\;11,\;36,\;\text{~and~} 47$~MeV. The interpolated eEST separable representation (eEST(intp)-c) is
   shown by a (blue) dash-dot-dotted line. The corresponding relative error is shown
   by (blue) crosses. Cubic splines were employed for the interpolation of $\mathcal{M}^c(E)$ 
   on the grid $E_{lab}= 5, 11,\;36,\text{~and~} 47$~MeV. 
}
\label{fig6}
\end{figure}

Next we examine the on-shell properties for proton scattering from $^{208}$Pb,
and concentrate on the $l=0$, $j=1/2$ partial wave. 
To keep
the relative error defined in Eq.~(\ref{eq:res2}) below 2$\%$ in the energy range from
0 to 50~MeV it is necessary to employ a rank-5 representation of the CH89 optical potential
in the lower partial waves~\cite{Hlophe:2014soa} even in the eEST scheme.
In this case the error is dominated by numerical interpolation when calculating the separable
representation. 
To compare
the eEST scheme to its energy-independent counterpart, we artificially lower the accuracy of the 
separable expansion to rank-4. This leads to a relative error
that is dominated by the quality of the separable representation over a the energy range under
consideration.
In panel (a) of Fig.~\ref{fig6} the  $S$ matrix elements obtained from the CH89 potential
are depicted together with their  EST and eEST rank-4 separable representations in the
energy range from 0 to 50~MeV. The relative errors with respect to the CH89 result give a
more detailed insight and are shown for the two different 
schemes in panel (b) of the same figure as filled circles for eEST and upward triangles
for EST scheme. As already observed in Section~\ref{result:nA},
the eEST scheme yields a better  representation of the $S$ matrix between 20 and 35~MeV than
the EST representation. 
By increasing the rank to a rank-5 representation, 
both representations can be improved for this energy interval.

\subsection{Approximation to the Energy Dependence}
 
The matrix elements $\mathcal{U}^{e,sc}_{ij}(E)$ are evaluated according 
to Eq.~(\ref{eq:pform5}). Additional numerical work is required to compute
the Coulomb distorted short-ranged potential $U^{sc}(k',k,E)$ at each energy $E$.
This makes the implementation of the eEST separable representation
computationally more involved compared to the energy-independent scheme. 
In cases where calculating the potential $U^s(E)$
in the plane wave basis is already time consuming, employing the eEST scheme
may become prohibitively costly. Therefore, it is worthwhile exploring if the eEST can
be modified in such a way that the potential  $U^s(E)$ is calculated only at a specified
fixed set of energies. 

In general the energy dependence of optical potentials is smooth and thus 
one may think of interpolating $\mathcal{U}^{e,sc}(E)$ on the energy variable.
We tested such an interpolation scheme starting by adopting
the energies of the support points at which the potential  $U^s(E)$ is already calculated
as grid points for an interpolation with Cubic Hermite splines~\cite{Huber:1996td}.  
The $S$ matrix elements evaluated using the interpolated eEST scheme is shown by a dash-dot-dotted 
line in panel (a) of  Fig.~\ref{fig6}, while the relative error is indicated by crosses in 
panel (b). 
As the figure illustrates, using an
interpolation to approximate the energy dependence of the CH89 potential yields the
same relative error as the exact eEST calculation. This means,
that the EST support points can already provide a good interpolation grid.
However, in cases with a more intricate energy dependence or if the distance between support points
is much larger, it may turn out to be 
necessary to add a few more energy points. 

The use of an energy interpolation greatly reduces the numerical effort needed to 
evaluate $\mathcal{U}^{e,sc}(E)$ on the energy grid from 0 to 50~MeV. The 
gain in computation time decreases with the number of interpolation points but
increases with the density of the energy grid. In the calculations presented here,
there are four interpolation points and the energy
grid consists of 100 points. Employing an interpolation to approximate the matrix elements 
$\mathcal{U}^{e,sc}_{ij}(E)$  reduces the computational effort by a factor of 23.
As pointed out in Section~\ref{results:pA}, a more accurate separable representation
of the s-wave $S$ matrix for the p+$^{208}$Pb system requires five EST support points. For
this case there are five interpolation points and the computation time is decreased by 
a factor of 18.
 

\section{Summary and Conclusions}
\label{summary}

In this work we introduce an explicit energy dependence into our previously developed 
separable representation of two-body transition matrix elements as well as potentials for 
nucleon-nucleus~\cite{Hlophe:2013xca} and proton-nucleus~\cite{Hlophe:2014soa} phenomenological
global optical potentials. Those potentials are in general complex and energy-dependent.
While on-shell properties like scattering amplitudes and cross sections can be 
well reproduced by an energy-independent separable representation, for which we generalized the
Ernst-Shakin-Thaler~\cite{Ernst:1973zzb} (EST) scheme to complex
potentials~\cite{Hlophe:2013xca}, the so obtained 
fully-off shell separable transition matrix elements fulfill the reciprocity theorem only
approximately, specifically when going far off the energy shell. The reason for this behavior 
lies in the energy dependence of the optical potential for which we construct the separable
representation. Specifically, the conditions for the coupling constants posed by the EST
construction, Eqs.~(\ref{eq:form1b}), can not be fulfilled simultaneously when the potential is
energy dependent. This insight had already been pointed out in Ref.~\cite{Pearce:1987zz} and
the EST scheme was corrected for Hermitian energy-dependent potentials. Picking up these
suggestions and applying them to non-Hermitian, energy-dependent optical potentials 
leads to an explicitly energy-dependent separable expansion, eEST, which fulfills reciprocity
exactly.  

As specific examples we consider neutron scattering from $^{48}$Ca and $^{208}$Pb described by
the Chapel Hill global phenomenological optical potential~\cite{Varner:1991zz}, and demonstrate 
for two different partial wave channels that the fully off-shell
transition matrix is exactly symmetric under the exchange of the off-shell momenta $k'$ and $k$,
and thus fulfills reciprocity. For the on-shell condition, we show that an energy-dependent
eEST representation of a partial wave $S$ matrix is slightly superior to its energy-independent
EST counterpart. However, one needs to note that any  separable representation can always be
improved on-shell by increasing its rank, while the symmetry property of the off-shell
transition amplitude is not affected by the rank. 

Since in a (d,p) reaction calculation one needs the effective interactions in the
neutron-nucleus as well as the proton-nucleus subsystem, we extended the energy-independent  EST
representation from Ref.~\cite{Hlophe:2014soa} to an energy-dependent one. The separable
representation of the proton-nucleus transition elements is carried out in the basis of Coulomb
scattering states. The calculation of required potential matrix elements follows the
approach suggested in Refs.~\cite{Elster:1993dv,Chinn:1991jb}. As test case we presented
the separable, energy-dependent representation of an $l=0$ partial wave off-shell transition
amplitude for proton scattering off $^{208}$Pb and demonstrated that it is exactly symmetric
under the exchange of the off-shell momenta $k'$ and $k$, thus thus fulfills reciprocity.
Similar to the neutron case we also show that the energy-dependent separable representation of
the corresponding $S$ matrix elements are slightly superior to the energy-independent
representation for the same rank.

The numerical evaluation of an energy-dependent separable representation is more involved
compared to its energy-independent counterpart, since the coupling matrices are now energy
dependent and thus need to be evaluated at each energy, not only at the EST support points. 
Though this is not a particular issue for the phenomenological global optical potentials, it
may become computationally expensive for microscopic optical potentials. 
Therefore we investigated if it is possible to interpolate on the energy variable, which
usually exhibits a relatively smooth behavior for optical potentials. We found that, when
using the EST support points as interpolation points for an interpolation on the energy with
cubic splines, we obtained a separable, energy-dependent representation of identical quality,
while considerably reducing the computational effort. Even for cases exhibiting a strong energy
dependence of the potential, it will be possible to use an energy interpolation by making the
energy grid finer. 

Summarizing, by constructing an energy-dependent separable representation of neutron- and
proton-optical potentials, one can obtain off-shell transition matrix elements which fulfill
the reciprocity theorem exactly. Since off-shell matrix elements are not observables, only
reaction calculations can show how severe any consequences, e.g. for three-body observables,
small violations of the reciprocity theorem turn out to be.


\begin{acknowledgments}
This work was performed in part under the
auspices of the U.~S.  Department of Energy under contract
No. DE-FG02-93ER40756 with Ohio University.
The authors thank F.M. Nunes and I.J. Thompson
for thoughtful comments and careful reading of the manuscript.
\end{acknowledgments}


\bibliography{coulomb}

\begin{thebibliography}{24}%
\makeatletter
\providecommand \@ifxundefined [1]{%
 \@ifx{#1\undefined}
}%
\providecommand \@ifnum [1]{%
 \ifnum #1\expandafter \@firstoftwo
 \else \expandafter \@secondoftwo
 \fi
}%
\providecommand \@ifx [1]{%
 \ifx #1\expandafter \@firstoftwo
 \else \expandafter \@secondoftwo
 \fi
}%
\providecommand \natexlab [1]{#1}%
\providecommand \enquote  [1]{``#1''}%
\providecommand \bibnamefont  [1]{#1}%
\providecommand \bibfnamefont [1]{#1}%
\providecommand \citenamefont [1]{#1}%
\providecommand \href@noop [0]{\@secondoftwo}%
\providecommand \href [0]{\begingroup \@sanitize@url \@href}%
\providecommand \@href[1]{\@@startlink{#1}\@@href}%
\providecommand \@@href[1]{\endgroup#1\@@endlink}%
\providecommand \@sanitize@url [0]{\catcode `\\12\catcode `\$12\catcode
  `\&12\catcode `\#12\catcode `\^12\catcode `\_12\catcode `\%12\relax}%
\providecommand \@@startlink[1]{}%
\providecommand \@@endlink[0]{}%
\providecommand \url  [0]{\begingroup\@sanitize@url \@url }%
\providecommand \@url [1]{\endgroup\@href {#1}{\urlprefix }}%
\providecommand \urlprefix  [0]{URL }%
\providecommand \Eprint [0]{\href }%
\providecommand \doibase [0]{http://dx.doi.org/}%
\providecommand \selectlanguage [0]{\@gobble}%
\providecommand \bibinfo  [0]{\@secondoftwo}%
\providecommand \bibfield  [0]{\@secondoftwo}%
\providecommand \translation [1]{[#1]}%
\providecommand \BibitemOpen [0]{}%
\providecommand \bibitemStop [0]{}%
\providecommand \bibitemNoStop [0]{.\EOS\space}%
\providecommand \EOS [0]{\spacefactor3000\relax}%
\providecommand \BibitemShut  [1]{\csname bibitem#1\endcsname}%
\let\auto@bib@innerbib\@empty
\bibitem [{\citenamefont {Kozub}\ \emph {et~al.}(2012)\citenamefont {Kozub},
  \citenamefont {Arbanas}, \citenamefont {Adekola}, \citenamefont {Bardayan},
  \citenamefont {Blackmon} \emph {et~al.}}]{Kozub:2012ka}%
  \BibitemOpen
  \bibfield  {author} {\bibinfo {author} {\bibfnamefont {R.}~\bibnamefont
  {Kozub}}, \bibinfo {author} {\bibfnamefont {G.}~\bibnamefont {Arbanas}},
  \bibinfo {author} {\bibfnamefont {A.}~\bibnamefont {Adekola}}, \bibinfo
  {author} {\bibfnamefont {D.}~\bibnamefont {Bardayan}}, \bibinfo {author}
  {\bibfnamefont {J.}~\bibnamefont {Blackmon}},  \emph {et~al.},\ }\href
  {\doibase 10.1103/PhysRevLett.109.172501} {\bibfield  {journal} {\bibinfo
  {journal} {Phys.Rev.Lett.}\ }\textbf {\bibinfo {volume} {109}},\ \bibinfo
  {pages} {172501} (\bibinfo {year} {2012})}\BibitemShut {NoStop}%
\bibitem [{\citenamefont {Thompson}\ and\ \citenamefont
  {Nunes}(2009)}]{ThompsonNunes}%
  \BibitemOpen
  \bibfield  {author} {\bibinfo {author} {\bibfnamefont {I.~J.}\ \bibnamefont
  {Thompson}}\ and\ \bibinfo {author} {\bibfnamefont {F.~M.}\ \bibnamefont
  {Nunes}},\ }\href@noop {} {\emph {\bibinfo {title} {{Nuclear Reactions for
  Astrophysics}}}}\ (\bibinfo  {publisher} {Cambridge University Press},\
  \bibinfo {year} {2009})\BibitemShut {NoStop}%
\bibitem [{\citenamefont {Alt}\ \emph {et~al.}(1967)\citenamefont {Alt},
  \citenamefont {Grassberger},\ and\ \citenamefont {Sandhas}}]{ags}%
  \BibitemOpen
  \bibfield  {author} {\bibinfo {author} {\bibfnamefont {E.}~\bibnamefont
  {Alt}}, \bibinfo {author} {\bibfnamefont {P.}~\bibnamefont {Grassberger}}, \
  and\ \bibinfo {author} {\bibfnamefont {W.}~\bibnamefont {Sandhas}},\
  }\href@noop {} {\bibfield  {journal} {\bibinfo  {journal} {Nucl. Phys. B}\
  }\textbf {\bibinfo {volume} {2}},\ \bibinfo {pages} {167} (\bibinfo {year}
  {1967})}\BibitemShut {NoStop}%
\bibitem [{\citenamefont {Deltuva}\ and\ \citenamefont
  {Fonseca}(2009)}]{Deltuva:2009fp}%
  \BibitemOpen
  \bibfield  {author} {\bibinfo {author} {\bibfnamefont {A.}~\bibnamefont
  {Deltuva}}\ and\ \bibinfo {author} {\bibfnamefont {A.}~\bibnamefont
  {Fonseca}},\ }\href {\doibase 10.1103/PhysRevC.79.014606} {\bibfield
  {journal} {\bibinfo  {journal} {Phys.Rev.}\ }\textbf {\bibinfo {volume}
  {C79}},\ \bibinfo {pages} {014606} (\bibinfo {year} {2009})}\BibitemShut
  {NoStop}%
\bibitem [{\citenamefont {Deltuva}\ \emph
  {et~al.}(2005{\natexlab{a}})\citenamefont {Deltuva}, \citenamefont
  {Fonseca},\ and\ \citenamefont {Sauer}}]{Deltuva:2005wx}%
  \BibitemOpen
  \bibfield  {author} {\bibinfo {author} {\bibfnamefont {A.}~\bibnamefont
  {Deltuva}}, \bibinfo {author} {\bibfnamefont {A.}~\bibnamefont {Fonseca}}, \
  and\ \bibinfo {author} {\bibfnamefont {P.}~\bibnamefont {Sauer}},\ }\href
  {\doibase 10.1103/PhysRevC.71.054005} {\bibfield  {journal} {\bibinfo
  {journal} {Phys.Rev.}\ }\textbf {\bibinfo {volume} {C71}},\ \bibinfo {pages}
  {054005} (\bibinfo {year} {2005}{\natexlab{a}})}\BibitemShut {NoStop}%
\bibitem [{\citenamefont {Deltuva}\ \emph
  {et~al.}(2005{\natexlab{b}})\citenamefont {Deltuva}, \citenamefont
  {Fonseca},\ and\ \citenamefont {Sauer}}]{Deltuva:2005cc}%
  \BibitemOpen
  \bibfield  {author} {\bibinfo {author} {\bibfnamefont {A.}~\bibnamefont
  {Deltuva}}, \bibinfo {author} {\bibfnamefont {A.}~\bibnamefont {Fonseca}}, \
  and\ \bibinfo {author} {\bibfnamefont {P.}~\bibnamefont {Sauer}},\ }\href
  {\doibase 10.1103/PhysRevC.72.059903, 10.1103/PhysRevC.72.054004} {\bibfield
  {journal} {\bibinfo  {journal} {Phys.Rev.}\ }\textbf {\bibinfo {volume}
  {C72}},\ \bibinfo {pages} {054004} (\bibinfo {year}
  {2005}{\natexlab{b}})}\BibitemShut {NoStop}%
\bibitem [{\citenamefont {Nunes}\ and\ \citenamefont
  {Upadhyay}(2012)}]{hites-proc}%
  \BibitemOpen
  \bibfield  {author} {\bibinfo {author} {\bibfnamefont {F.}~\bibnamefont
  {Nunes}}\ and\ \bibinfo {author} {\bibfnamefont {N.}~\bibnamefont
  {Upadhyay}},\ }\href {\doibase doi:10.1088/1742-6596/403/1/012029} {\bibfield
   {journal} {\bibinfo  {journal} {J. Phys. G: Conf. Ser.}\ }\textbf {\bibinfo
  {volume} {403}},\ \bibinfo {pages} {012029} (\bibinfo {year}
  {2012})}\BibitemShut {NoStop}%
\bibitem [{\citenamefont {Mukhamedzhanov}\ \emph {et~al.}(2012)\citenamefont
  {Mukhamedzhanov}, \citenamefont {Eremenko},\ and\ \citenamefont
  {Sattarov}}]{Mukhamedzhanov:2012qv}%
  \BibitemOpen
  \bibfield  {author} {\bibinfo {author} {\bibfnamefont {A.}~\bibnamefont
  {Mukhamedzhanov}}, \bibinfo {author} {\bibfnamefont {V.}~\bibnamefont
  {Eremenko}}, \ and\ \bibinfo {author} {\bibfnamefont {A.}~\bibnamefont
  {Sattarov}},\ }\href {\doibase 10.1103/PhysRevC.86.034001} {\bibfield
  {journal} {\bibinfo  {journal} {Phys.Rev.}\ }\textbf {\bibinfo {volume}
  {C86}},\ \bibinfo {pages} {034001} (\bibinfo {year} {2012})}\BibitemShut
  {NoStop}%
\bibitem [{\citenamefont {Upadhyay}\ \emph {et~al.}(2014)\citenamefont
  {Upadhyay} \emph {et~al.}}]{upadhyay:2014}%
  \BibitemOpen
  \bibfield  {author} {\bibinfo {author} {\bibfnamefont {N.}~\bibnamefont
  {Upadhyay}} \emph {et~al.} (\bibinfo {collaboration} {TORUS Collaboration}),\
  }\href@noop {} {\bibfield  {journal} {\bibinfo  {journal} {Phys. Rev.}\
  }\textbf {\bibinfo {volume} {C90}},\ \bibinfo {pages} {014615} (\bibinfo
  {year} {2014})}\BibitemShut {NoStop}%
\bibitem [{\citenamefont {Haidenbauer}\ and\ \citenamefont
  {Plessas}(1983)}]{Haidenbauer:1982if}%
  \BibitemOpen
  \bibfield  {author} {\bibinfo {author} {\bibfnamefont {J.}~\bibnamefont
  {Haidenbauer}}\ and\ \bibinfo {author} {\bibfnamefont {W.}~\bibnamefont
  {Plessas}},\ }\href {\doibase 10.1103/PhysRevC.27.63} {\bibfield  {journal}
  {\bibinfo  {journal} {Phys.Rev.}\ }\textbf {\bibinfo {volume} {C27}},\
  \bibinfo {pages} {63} (\bibinfo {year} {1983})}\BibitemShut {NoStop}%
\bibitem [{\citenamefont {Haidenbauer}\ \emph {et~al.}(1986)\citenamefont
  {Haidenbauer}, \citenamefont {Koike},\ and\ \citenamefont
  {Plessas}}]{Haidenbauer:1986zza}%
  \BibitemOpen
  \bibfield  {author} {\bibinfo {author} {\bibfnamefont {J.}~\bibnamefont
  {Haidenbauer}}, \bibinfo {author} {\bibfnamefont {Y.}~\bibnamefont {Koike}},
  \ and\ \bibinfo {author} {\bibfnamefont {W.}~\bibnamefont {Plessas}},\ }\href
  {\doibase 10.1103/PhysRevC.33.439} {\bibfield  {journal} {\bibinfo  {journal}
  {Phys.Rev.}\ }\textbf {\bibinfo {volume} {C33}},\ \bibinfo {pages} {439}
  (\bibinfo {year} {1986})}\BibitemShut {NoStop}%
\bibitem [{\citenamefont {Entem}\ \emph {et~al.}(2001)\citenamefont {Entem},
  \citenamefont {Fernandez},\ and\ \citenamefont {Valcarce}}]{Entem:2001it}%
  \BibitemOpen
  \bibfield  {author} {\bibinfo {author} {\bibfnamefont {D.}~\bibnamefont
  {Entem}}, \bibinfo {author} {\bibfnamefont {F.}~\bibnamefont {Fernandez}}, \
  and\ \bibinfo {author} {\bibfnamefont {A.}~\bibnamefont {Valcarce}},\ }\href
  {\doibase 10.1088/0954-3899/27/7/312} {\bibfield  {journal} {\bibinfo
  {journal} {J.Phys.}\ }\textbf {\bibinfo {volume} {G27}},\ \bibinfo {pages}
  {1537} (\bibinfo {year} {2001})}\BibitemShut {NoStop}%
\bibitem [{\citenamefont {Ueda}\ and\ \citenamefont
  {Ikegami}(1994)}]{Ueda:1994ur}%
  \BibitemOpen
  \bibfield  {author} {\bibinfo {author} {\bibfnamefont {T.}~\bibnamefont
  {Ueda}}\ and\ \bibinfo {author} {\bibfnamefont {Y.}~\bibnamefont {Ikegami}},\
  }\href {\doibase 10.1143/PTP.91.85} {\bibfield  {journal} {\bibinfo
  {journal} {Prog.Theor.Phys.}\ }\textbf {\bibinfo {volume} {91}},\ \bibinfo
  {pages} {85} (\bibinfo {year} {1994})}\BibitemShut {NoStop}%
\bibitem [{\citenamefont {Gal}\ and\ \citenamefont
  {Garcilazo}(2011)}]{Gal:2011yp}%
  \BibitemOpen
  \bibfield  {author} {\bibinfo {author} {\bibfnamefont {A.}~\bibnamefont
  {Gal}}\ and\ \bibinfo {author} {\bibfnamefont {H.}~\bibnamefont
  {Garcilazo}},\ }\href {\doibase 10.1016/j.nuclphysa.2011.06.022} {\bibfield
  {journal} {\bibinfo  {journal} {Nucl.Phys.}\ }\textbf {\bibinfo {volume}
  {A864}},\ \bibinfo {pages} {153} (\bibinfo {year} {2011})}\BibitemShut
  {NoStop}%
\bibitem [{\citenamefont {Ernst}\ \emph {et~al.}(1973)\citenamefont {Ernst},
  \citenamefont {Shakin},\ and\ \citenamefont {Thaler}}]{Ernst:1973zzb}%
  \BibitemOpen
  \bibfield  {author} {\bibinfo {author} {\bibfnamefont {D.~J.}\ \bibnamefont
  {Ernst}}, \bibinfo {author} {\bibfnamefont {C.~M.}\ \bibnamefont {Shakin}}, \
  and\ \bibinfo {author} {\bibfnamefont {R.~M.}\ \bibnamefont {Thaler}},\
  }\href {\doibase 10.1103/PhysRevC.8.46} {\bibfield  {journal} {\bibinfo
  {journal} {Phys.Rev.}\ }\textbf {\bibinfo {volume} {C8}},\ \bibinfo {pages}
  {46} (\bibinfo {year} {1973})}\BibitemShut {NoStop}%
\bibitem [{\citenamefont {Hlophe}\ \emph {et~al.}(2013)\citenamefont {Hlophe}
  \emph {et~al.}}]{Hlophe:2013xca}%
  \BibitemOpen
  \bibfield  {author} {\bibinfo {author} {\bibfnamefont {L.}~\bibnamefont
  {Hlophe}} \emph {et~al.} (\bibinfo {collaboration} {The TORUS
  Collaboration}),\ }\href {\doibase 10.1103/PhysRevC.88.064608} {\bibfield
  {journal} {\bibinfo  {journal} {Phys.Rev.}\ }\textbf {\bibinfo {volume}
  {C88}},\ \bibinfo {pages} {064608} (\bibinfo {year} {2013})},\ \Eprint
  {http://arxiv.org/abs/1310.8334} {arXiv:1310.8334 [nucl-th]} \BibitemShut
  {NoStop}%
\bibitem [{\citenamefont {Hlophe}\ \emph {et~al.}(2014)\citenamefont {Hlophe},
  \citenamefont {Eremenko}, \citenamefont {Elster}, \citenamefont {Nunes},
  \citenamefont {Arbanas}, \citenamefont {Escher},\ and\ \citenamefont
  {Thompson}}]{Hlophe:2014soa}%
  \BibitemOpen
  \bibfield  {author} {\bibinfo {author} {\bibfnamefont {L.}~\bibnamefont
  {Hlophe}}, \bibinfo {author} {\bibfnamefont {V.}~\bibnamefont {Eremenko}},
  \bibinfo {author} {\bibfnamefont {C.}~\bibnamefont {Elster}}, \bibinfo
  {author} {\bibfnamefont {F.~M.}\ \bibnamefont {Nunes}}, \bibinfo {author}
  {\bibfnamefont {G.}~\bibnamefont {Arbanas}}, \bibinfo {author} {\bibfnamefont
  {J.~E.}\ \bibnamefont {Escher}}, \ and\ \bibinfo {author} {\bibfnamefont
  {I.~J.}\ \bibnamefont {Thompson}},\ }\href {\doibase
  10.1103/PhysRevC.90.061602} {\bibfield  {journal} {\bibinfo  {journal} {Phys.
  Rev.}\ }\textbf {\bibinfo {volume} {C90}},\ \bibinfo {pages} {061602}
  (\bibinfo {year} {2014})},\ \Eprint {http://arxiv.org/abs/1409.4012}
  {arXiv:1409.4012 [nucl-th]} \BibitemShut {NoStop}%
\bibitem [{\citenamefont {Pearce}(1987)}]{Pearce:1987zz}%
  \BibitemOpen
  \bibfield  {author} {\bibinfo {author} {\bibfnamefont {B.}~\bibnamefont
  {Pearce}},\ }\href {\doibase 10.1103/PhysRevC.36.471} {\bibfield  {journal}
  {\bibinfo  {journal} {Phys.Rev.}\ }\textbf {\bibinfo {volume} {C36}},\
  \bibinfo {pages} {471} (\bibinfo {year} {1987})}\BibitemShut {NoStop}%
\bibitem [{\citenamefont {Pearce}\ and\ \citenamefont
  {Afnan}(1989)}]{Pearce:1989re}%
  \BibitemOpen
  \bibfield  {author} {\bibinfo {author} {\bibfnamefont {B.~C.}\ \bibnamefont
  {Pearce}}\ and\ \bibinfo {author} {\bibfnamefont {I.~R.}\ \bibnamefont
  {Afnan}},\ }\href {\doibase 10.1103/PhysRevC.40.220} {\bibfield  {journal}
  {\bibinfo  {journal} {Phys. Rev.}\ }\textbf {\bibinfo {volume} {C40}},\
  \bibinfo {pages} {220} (\bibinfo {year} {1989})}\BibitemShut {NoStop}%
\bibitem [{\citenamefont {Saito}\ and\ \citenamefont
  {Haidenbauer}(2000)}]{Saito:2000bg}%
  \BibitemOpen
  \bibfield  {author} {\bibinfo {author} {\bibfnamefont {T.~Y.}\ \bibnamefont
  {Saito}}\ and\ \bibinfo {author} {\bibfnamefont {J.}~\bibnamefont
  {Haidenbauer}},\ }\href {\doibase 10.1007/s100500050428, 10.1007/PL00013657}
  {\bibfield  {journal} {\bibinfo  {journal} {Eur. Phys. J.}\ }\textbf
  {\bibinfo {volume} {A7}},\ \bibinfo {pages} {559} (\bibinfo {year} {2000})},\
  \Eprint {http://arxiv.org/abs/nucl-th/0003064} {arXiv:nucl-th/0003064
  [nucl-th]} \BibitemShut {NoStop}%
\bibitem [{\citenamefont {Varner}\ \emph {et~al.}(1991)\citenamefont {Varner},
  \citenamefont {Thompson}, \citenamefont {McAbee}, \citenamefont {Ludwig},\
  and\ \citenamefont {Clegg}}]{Varner:1991zz}%
  \BibitemOpen
  \bibfield  {author} {\bibinfo {author} {\bibfnamefont {R.}~\bibnamefont
  {Varner}}, \bibinfo {author} {\bibfnamefont {W.}~\bibnamefont {Thompson}},
  \bibinfo {author} {\bibfnamefont {T.}~\bibnamefont {McAbee}}, \bibinfo
  {author} {\bibfnamefont {E.}~\bibnamefont {Ludwig}}, \ and\ \bibinfo {author}
  {\bibfnamefont {T.}~\bibnamefont {Clegg}},\ }\href {\doibase 10.10
  16/0370-1573(91)90039-O} {\bibfield  {journal} {\bibinfo  {journal}
  {Phys.Rept.}\ }\textbf {\bibinfo {volume} {201}},\ \bibinfo {pages} {57}
  (\bibinfo {year} {1991})}\BibitemShut {NoStop}%
\bibitem [{\citenamefont {Elster}\ \emph {et~al.}(1993)\citenamefont {Elster},
  \citenamefont {Liu},\ and\ \citenamefont {Thaler}}]{Elster:1993dv}%
  \BibitemOpen
  \bibfield  {author} {\bibinfo {author} {\bibfnamefont {C.}~\bibnamefont
  {Elster}}, \bibinfo {author} {\bibfnamefont {L.~C.}\ \bibnamefont {Liu}}, \
  and\ \bibinfo {author} {\bibfnamefont {R.~M.}\ \bibnamefont {Thaler}},\
  }\href {\doibase 10.1088/0954-3899/19/12/015} {\bibfield  {journal} {\bibinfo
   {journal} {J.Phys.}\ }\textbf {\bibinfo {volume} {G19}},\ \bibinfo {pages}
  {2123} (\bibinfo {year} {1993})}\BibitemShut {NoStop}%
\bibitem [{\citenamefont {Chinn}\ \emph {et~al.}(1991)\citenamefont {Chinn},
  \citenamefont {Elster},\ and\ \citenamefont {Thaler}}]{Chinn:1991jb}%
  \BibitemOpen
  \bibfield  {author} {\bibinfo {author} {\bibfnamefont {C.~R.}\ \bibnamefont
  {Chinn}}, \bibinfo {author} {\bibfnamefont {C.}~\bibnamefont {Elster}}, \
  and\ \bibinfo {author} {\bibfnamefont {R.~M.}\ \bibnamefont {Thaler}},\
  }\href {\doibase 10.1103/PhysRevC.44.1569} {\bibfield  {journal} {\bibinfo
  {journal} {Phys.Rev.}\ }\textbf {\bibinfo {volume} {C44}},\ \bibinfo {pages}
  {1569} (\bibinfo {year} {1991})}\BibitemShut {NoStop}%
\bibitem [{\citenamefont {Huber}\ \emph {et~al.}(1997)\citenamefont {Huber},
  \citenamefont {Witala}, \citenamefont {Nogga}, \citenamefont {Gloeckle},\
  and\ \citenamefont {Kamada}}]{Huber:1996td}%
  \BibitemOpen
  \bibfield  {author} {\bibinfo {author} {\bibfnamefont {D.}~\bibnamefont
  {Huber}}, \bibinfo {author} {\bibfnamefont {H.}~\bibnamefont {Witala}},
  \bibinfo {author} {\bibfnamefont {A.}~\bibnamefont {Nogga}}, \bibinfo
  {author} {\bibfnamefont {W.}~\bibnamefont {Gloeckle}}, \ and\ \bibinfo
  {author} {\bibfnamefont {H.}~\bibnamefont {Kamada}},\ }\href {\doibase
  10.1007/s006010050057} {\bibfield  {journal} {\bibinfo  {journal} {Few Body
  Syst.}\ }\textbf {\bibinfo {volume} {22}},\ \bibinfo {pages} {107} (\bibinfo
  {year} {1997})}\BibitemShut {NoStop}%
\end{thebibliography}%

\clearpage

\newpage

\noindent

\noindent

\end{document}